# Deep Learning Cell Imaging through Anderson Localizing Optical Fibre

Jian Zhao[1,*], Yangyang Sun[1], Hongbo Zhu[2], Zheyuan Zhu[1], Jose Enrique Antonio-Lopez[1], Rodrigo Amezcua Correa[1], Shuo Pang[1], & Axel Schülzgen[1]

[1]CREOL, College of Optics and Photonics, University of Central Florida, Orlando, FL 32816, USA
[2]State Key Laboratory of Luminescence and Applications, Changchun Institute of Optics, Fine Mechanics and Physics, Chinese Academy of Sciences, Changchun 130033, China
*JianZHAO@knights.ucf.edu

**ABSTRACT**

*We demonstrate a deep-learning-based fibre imaging system which can transfer real-time artifact-free cell images through a meter-long Anderson localizing optical fibre. The cell samples are illuminated by an incoherent LED light source. A deep convolutional neural network is applied to the image reconstruction process. The network training uses data generated by a set-up with straight fibre at room temperature (~20 °C) but can be utilized directly for high fidelity reconstruction of cell images that are transported through fibre with a few degrees bend and/or fibre with segments heated up to 50 °C. In addition, cell images located several millimeters away from the bare fibre end can be transported and recovered successfully without the assistance of any distal optics. We further evidence that the trained neural network is able to reconstruct the images of cells which are never used in the training process and feature very different morphology.*

In neuroscience and clinical applications, visualizations of real-time cell activity, morphology and overall tissue architecture are crucial for fundamental research and medical diagnosis[1-4]. This usually requires real-time *in vivo* imaging to be performed in a minimally invasive way with the ability to deeply penetrate into organs. Due to the miniature size and flexible imaging transfer capability, fibre-optic imaging systems (FOISs) have been widely applied to this domain[4-13]. Current solutions are faced with challenges regarding bulky and complex distal optics, imaging artifacts and extreme sensitivity to perturbations. These limitations mainly originate from the optical fibre device. For example, multicore fibre bundles and multimode fibres (MMFs) are the two most widely used fibres in these systems. Systems based on fibre bundles usually require extra distal optics or mechanical actuators which limiting the extent of miniaturization[3,5,11]. In addition, the particular core patterns featured in fibre bundles result in pixelated artifacts in transported images[4,14-17]. Typical systems using MMF rely on image reconstruction processes using the transmission matrix method to compensate for randomized phases through wavefront shaping[8,9,13,18]. This kind of reconstruction process is vulnerable to perturbations. Even minor changes of temperature (a few degrees Celsius) or slight fibre movement (a few hundred micrometers) can induce mode coupling and scramble the pre-calibrated transmission matrix[13].

Recent burgeoning deep learning technology opens a new avenue for overcoming the physical limitation of fibre device[19]. Deep learning technology is a fast-developing research field which has gained great success in imaging applications and demonstrated better performance than conventional model-based methods[20-24]. Instead of relying on known models and priors, the deep convolutional neural network (DCNN) directly learns the underlying physics of imaging transmission system through a training process using a large dataset without any advanced knowledge. Especially, when it is difficult to develop an accurate physics model, this learning capability demonstrates its superiority over model-based methods. The trained DCNN is a precise approximation of the mapping function between the measured imaging data and the input imaging data. Well designed and trained DCNNs can be used to predict input images even if the particular type of images is not included in the set of training data. The prediction process usually takes less than one second on a regular GPU. Recently, the use of DCNNs for image recovery and classification after transport through optical fibres has been reported[25-28]. This combination opens new avenues for tremendous improvements of FOISs. For image transmission, two different types of optical fibres, MMF[25-28] and glass-air Anderson localizing optical fibre (GALOF)[25-28], have been utilized in recently reported DCNN-based FOISs. The DCNN-MMF based system is very sensitive to temperature variation and bending[25]. In contrast, the DCNN-GALOF system demonstrated bending-independent imaging capabilities[28]. This robust performance is based on the special mode properties of the GALOF[29-31]. The modes embedded in the random structure of GALOF are formed by multiple scattering process. Each mode corresponds to a beam transmission channel. The imaging information is encoded and transferred by thousands of modes in the GALOF. Unlike MMF, most of these modes mediated by transverse Anderson localization demonstrate single-mode properties, which makes the device rather insensitive to external perturbations. The numerical simulation of the wave propagation process inside such structure requires huge computational power[32]. Therefore, combining the deep learning with GALOF is a preferable scheme[28].

Nevertheless, the previous DCNN-GALOF fibre imaging system is faced with several challenges limiting its practical application. First, the system only demonstrated success in imaging of low-resolution sparse objects, such as the binary MNIST handwritten numbers. There is a chasm between spare objects reconstruction and the reconstruction of biological objects which are typically different types of cells or tissue with complicated morphologic features. In order to accomplish the demanding task to resolve and reconstruct the subtle details of a biological object, improved DCNNs which can "learn" the underlying physics of imaging process with higher precision are required. Second, the demonstrated transfer-learning capability of the previous DCNN-GALOF system was limited to binary sparse testing objects that were quite similar to the objects in the training data[27,28]. It would be more confirmative if the system would feature transfer learning performed using objects with features very different from the training data. Third, the illumination of the previous DCNN-GALOF system relies on high-brightness, coherent laser light sources. The coherence of lasers results in speckle patterns which reduce the image quality. In addition, the high intensity of laser light might be damaging to biological objects such as living cells and the cost of lasers is relatively high. Comparing to laser light source, incoherent illumination, such as LED illumination, can generally give better imaging quality without speckle patterns. The low intensity of incoherent light also helps protect cells against photobleaching and phototoxicity during the imaging process. At the same time, the cost of LEDs is much lower compared to laser systems[33]. Moreover, some medical practitioners might prefer incoherent light illumination. For example, white-light transmission cellular micrographs are already very familiar to histopathologists[3,11]. Therefore, to develop a system using incoherent illumination is an important step towards practical applications.

To resolve all abovementioned issues, we develop an incoherent light illuminated DCNN-GALOF imaging system with the capability to image various cell structures. Within this system, a new DCNN model with a tailored design is applied to the image reconstruction process, and a low-cost LED works as the light source. We call the new system Cell-DCNN-GALOF. We demonstrate that it is able to transfer high quality, artifact-free images of different types of cells in real time. We further prove that the imaging depth of this system can reach up to several millimeters without any distal optics. In addition, we show that the image reconstruction process is remarkable robust with regard to external perturbations, such as temperature variation and fibre bending. Last but not least, the transfer-learning capability of the new system is confirmed by using cells of different morphology for testing. The work presented here introduces a new platform for various practical applications, such as neuroscience research and clinical diagnosis. It is also a new cornerstone for imaging research based on waveguide devices using transverse Anderson localization.

**Results**
**Imaging of multiple cell types**

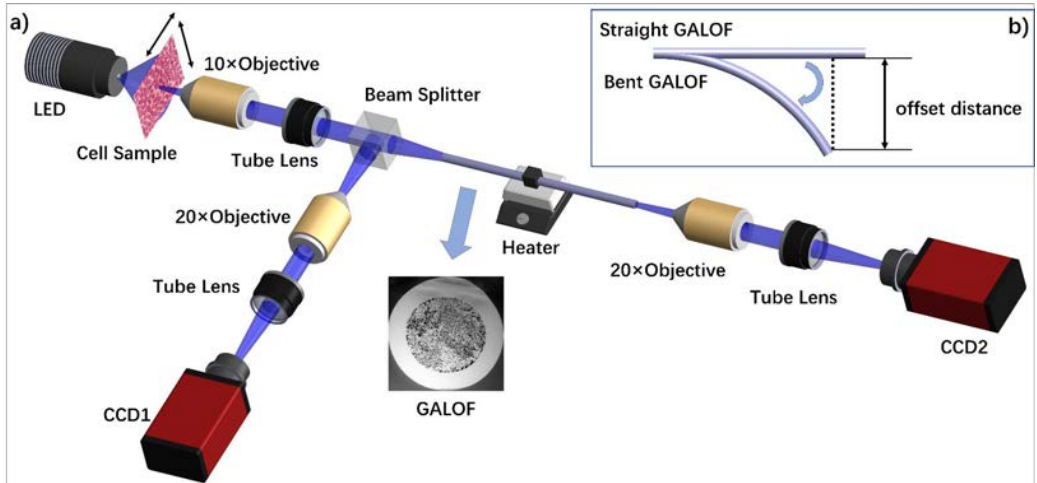

**Figure 1 | Schematic of the cell imaging setup. a)** The light source is a LED with a center wavelength of 460 nm. An 80 cm long GALOF sample is utilized. The SEM image of the GALOF cross-section is also shown in a). The diameter of the disordered structure is about 278 µm, and the air-hole-filling fraction in the disordered structure is approximately 28.5%. The temperature of a GALOF segment can be raised by the heater underneath. The images of cell samples are magnified by a 10x objective and split into two copies sent into a reference path and a measurement path, respectively. The cell samples are scanned both vertically and horizontally with 5 µm steps to obtain training, validation and test data sets. In the reference beam path, the image is further magnified by a 20x objective and recorded by CCD 1 (Manta G-145B) after passing through a tube lens. In the measurement path, the image is transported through the 80 cm long GALOF and then projected onto CCD 2 (Manta G-145B) by the same combination of 20x objective and tube lens. **b)** Experiments are performed for both straight GALOF and bent GALOF. To quantify the amount of bending, the offset distance is measured which is defined as the distance from the end of the bent fibre to the position of the straight fibre (equal to the length of the dashed line).

The experimental setup is shown in Fig. 1. Reference images and raw images of different areas of cell samples are recorded by CCD 1 and CCD 2, respectively, while the samples are being scanned. The reference images are labeled as the ground truth. Both reference and raw images are 8-bit grayscale images and are cropped to a size of 418x418 pixels. To demonstrate the imaging reconstruction capability, two different types of cells, human red blood cells and cancerous human stomach cells, serve as objects. By scanning across different areas of the cell sample, we collect 15000 reference and raw images as the training set, 1000 image pairs as the validation set, and another 1000 image pairs as the test set for each type of cell. During the first data acquisition process, the GALOF is kept straight and at room temperature of about 20° C. The imaging depth is 0 mm, meaning that the image plane is located directly at the fibre input facet. The training data are loaded into the DCNN (see Fig. 2 for DCNN structure) to optimize the parameters of the neural network and generate a computational architecture that can accurately map the fibre-transported images to the corresponding original object. After the training process, the test data are applied to the trained model to perform imaging reconstruction and evaluate its performance using the normalized mean absolute error (MAE) as the metric (see Methods). In the first round of experiments, we train and test each type of cell separately. With a training data set of 15000 image pairs, it takes about 6.4 hour to train the DCCN over 80 epochs on two GPUs (GeForce GTX 1080 Ti) using a personal computer. The accuracy improvement curves for both training and validation process over all 80 epochs are provided in Fig. S1 of the supplementary information. After training, the reconstruction time of a single test image is about 0.05 second. Fig. 3 shows some samples from the test data set. In a) to c), reference images, raw images, and recovered images of three in succession collected and reconstructed images of human red cells are shown, while in d) to f) three images of cancerous stomach cells are presented. Comparing the reference images with the reconstructed images, it is clear that the separately trained DCNNs are able to reconstruct images of both cell types remarkably well. The averaged normalized test MAEs are 0.024 and 0.027 for the human red blood cells and the cancerous human stomach cells, respectively, with standard deviations of 0.006 and 0.011. To further highlight the real-time imaging capability of our system, we visualize the test process for these two cell types in a video (see the supplementary movie). This real-time imaging capability is highly desirable for many practical applications, such as *in situ* morphologic examination of living tissues in their native context for pathology[3].

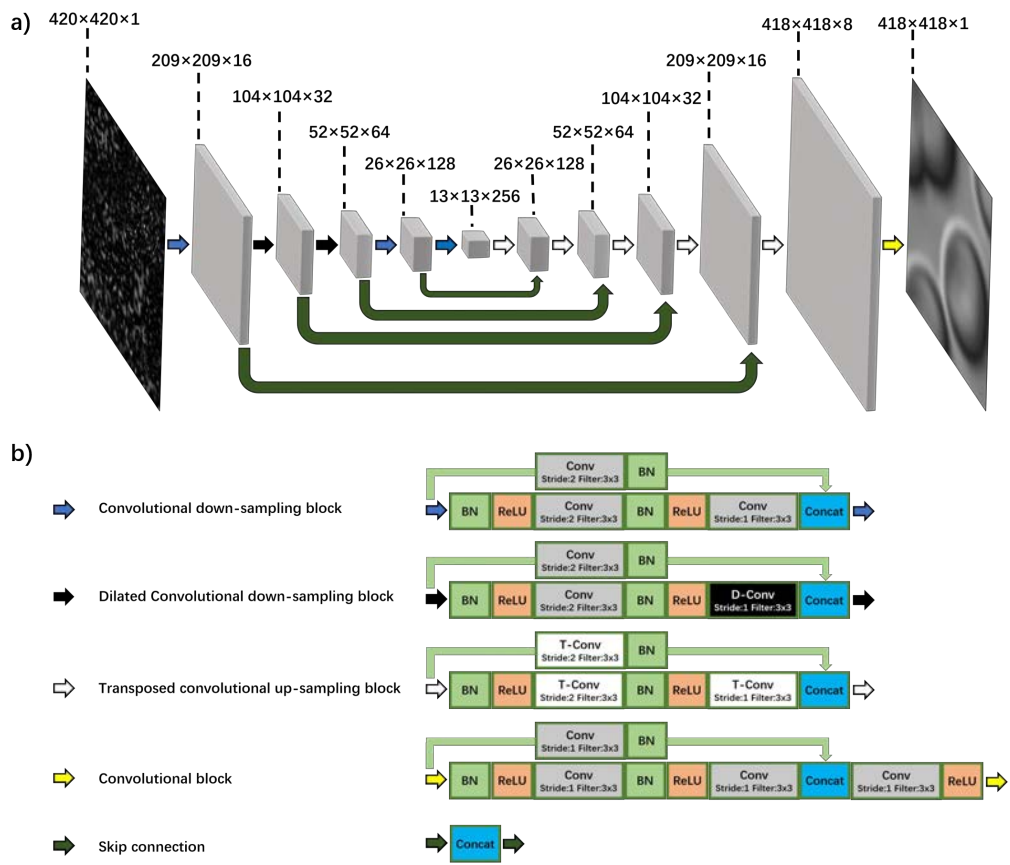

**Figure 2 | The architecture of the DCNN. a)** Detailed structure of the DCNN. The raw image which is resized to 420x420 using zero padding is the input layer. The input layer is decimated by five down-sampling blocks (blue and black arrows) to extract the feature maps. Then five up-sampling blocks (white arrows) and one convolutional block (yellow arrow) are applied to reconstruct the images of cell samples with a size of 418x418. The skip connections (dark green arrows) pass feature information from feature-extraction layers to reconstruction layers by concatenation operations. The MAE-based loss metrics are calculated by comparing the reconstructed images with the reference images. The parameters of the DCNN are optimized by minimizing the loss. **b)** Detailed block operation diagrams corresponding to the respective arrows shown on the right side (BN: Batch Normalization, ReLU: Rectified Linear Unit, Conv: Convolution, D-Conv: Dilated Convolution, T-Conv: Transposed Convolution, Concat: Concatenation).

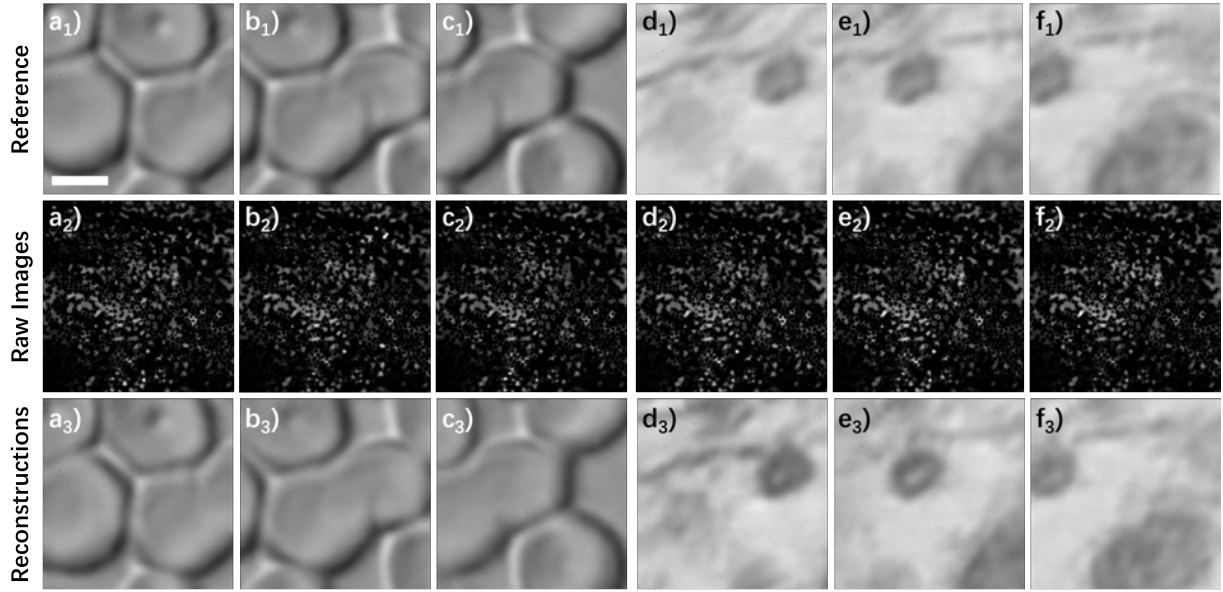

**Figure 3 | Cell imaging of different types of cells. a)-c)** are test data for human red blood cells. **d)-f)** are test data for cancerous human stomach cells. All data are collected with straight GALOF, at room temperature with 0 mm imaging depth. The length of the scale bar in **a$_1$)** is 4 μm. **a$_1$)-f$_1$)** are the reference images; **a$_2$)-f$_2$)** are the corresponding raw images. **a$_3$)-f$_3$)** are the images recovered from the raw images.

**Cell imaging at various depth**

Distal optics located at fibre input end hinders conventional FOIS from miniaturizing the size of the imaging unit. Here, we are investigating the ability of our Cell-DCNN-GALOF system to image objects located at various distances from the fibre input facet without distal optics. As illustrated in Fig. 4 g), the images of cells located at different imaging planes are collected by the bare fibre input end. The depth ranges from 0 mm to 5 mm with steps of 1 mm. For each individual depth, 15000 reference and raw images are collected as the training set, and another 1000 image pairs serve as the test set. The GALOF is kept straight and at room temperature during data collection. The DCNN is trained separately for each depth resulting in depth-specific parameters. Examining reference and reconstructed test images shown in Fig. 4 a) to f), high-quality image transmission and reconstruction can be achieved up to depths of at least 3 mm. The first visual degradation of the imaging quality appears around 4 mm and the visual quality of the reconstructed images drops further at 5 mm depth. The corresponding quantitative image quality evaluation is shown in Fig. 4 h). The normalized MAE increases almost linearly with a slope of about 0.008 per mm. Based on these data we conclude that our system can transfer high-quality cell images for objects being several mm away from the fibre input facet without the need for any distal optics. Therefore, the size of an image transmitting endoscope based on our system could be potentially minimized to the diameter of the fibre itself and the penetration damage could be reduced to a minimum without degrading the quality of the image of biological objects. The fibre could collect images of organs without touching them directly enabling a minimally invasive, high performance imaging system.

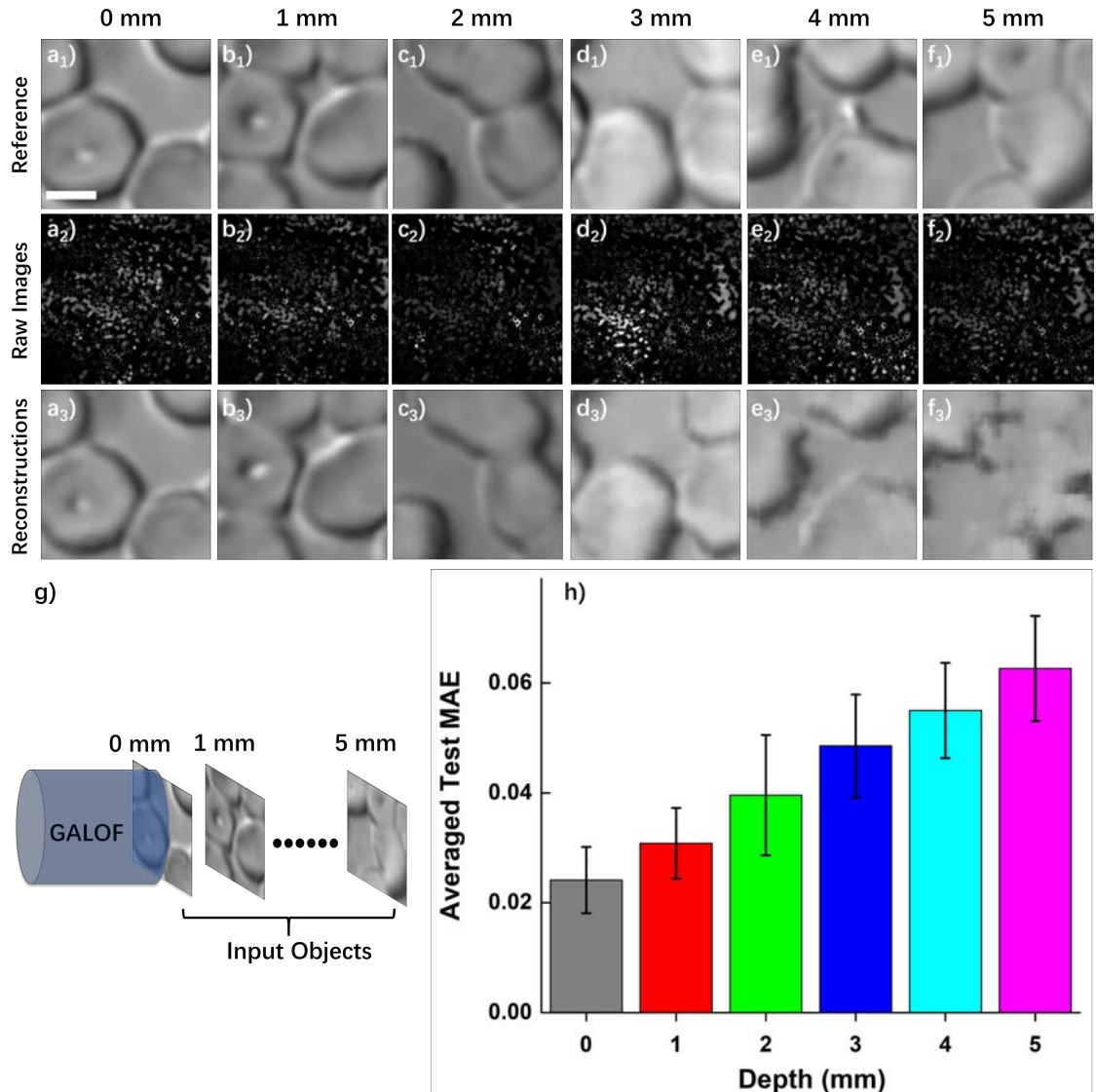

**Figure 4 | Multiple depth cell imaging. a)-f)** are test data for human red blood cells. All data are collected with straight GALOF at room temperature. All three images in each column are from the same depth. The length of the scale bar in **a$_1$)** is 4μm. **a$_1$)-f$_1$)** are the reference images; **a$_2$)-f$_2$)** are the corresponding raw images. As illustrated in **g)**, **a$_2$)-f$_2$)** are obtained by varying the imaging depth from 0 mm to 5 mm with steps of 1 mm. **a$_3$)-f$_3$)** are the images recovered from the corresponding raw images. **h)** is the averaged test MAE error for each depth with the standard deviation as the error bar. More sample results including reference, raw and recovered images are shown in Fig. S2 of the supplementary information.

**Cell imaging with temperature variation and fibre bending**

In practical applications, the optical fibre of the FOIS often needs to be inserted deeply into the cavity of living organs. This requires the imaging system to tolerate thermal variation and fibre bending. For MMF-based FOIS, the increase of temperature or bending of the fibre when inserting fibre into organs or tissues induces strong variations of the mode coupling. These variations decrease the performance of MMF-based imaging systems due to induced changes of the transmission matrix[13]. This problem can be overcome by using GALOF since most of the modes embedded in GALOF show single mode characteristics which increases the system tolerance and can make it immune even to rather strong perturbations. We first investigate the effect of temperature variation on our Cell-DCNN-GALOF system by changing the temperature of a 10 mm long GALOF segment with a heater. During the data collection, we keep the GALOF straight and at 0 mm imaging depth. We collect 15000 image pairs at 20 °C as the training data. For test data, we record three sets of test data where the GALOF segment is heated to 20 °C, 35 °C, and 50 °C, respectively. Each set of test data consists of 1000 image pairs. The DCNN model is only trained utilizing the training data collected at 20 °C. Subsequently, the trained model is applied to perform test image reconstruction of data acquired at all three different temperatures. In Fig. 5 a) to c), some sample images are shown. Comparing the reference with reconstructed images, the visual imaging quality is not affected by the thermal change even for a 30 °C variation. Most body temperatures of humans or animals fall into this range. This confirms the remarkable robustness of our Cell-DCNN-GALOF system regarding temperature fluctuations, which makes the system particularly suitable for *in vivo* imaging.

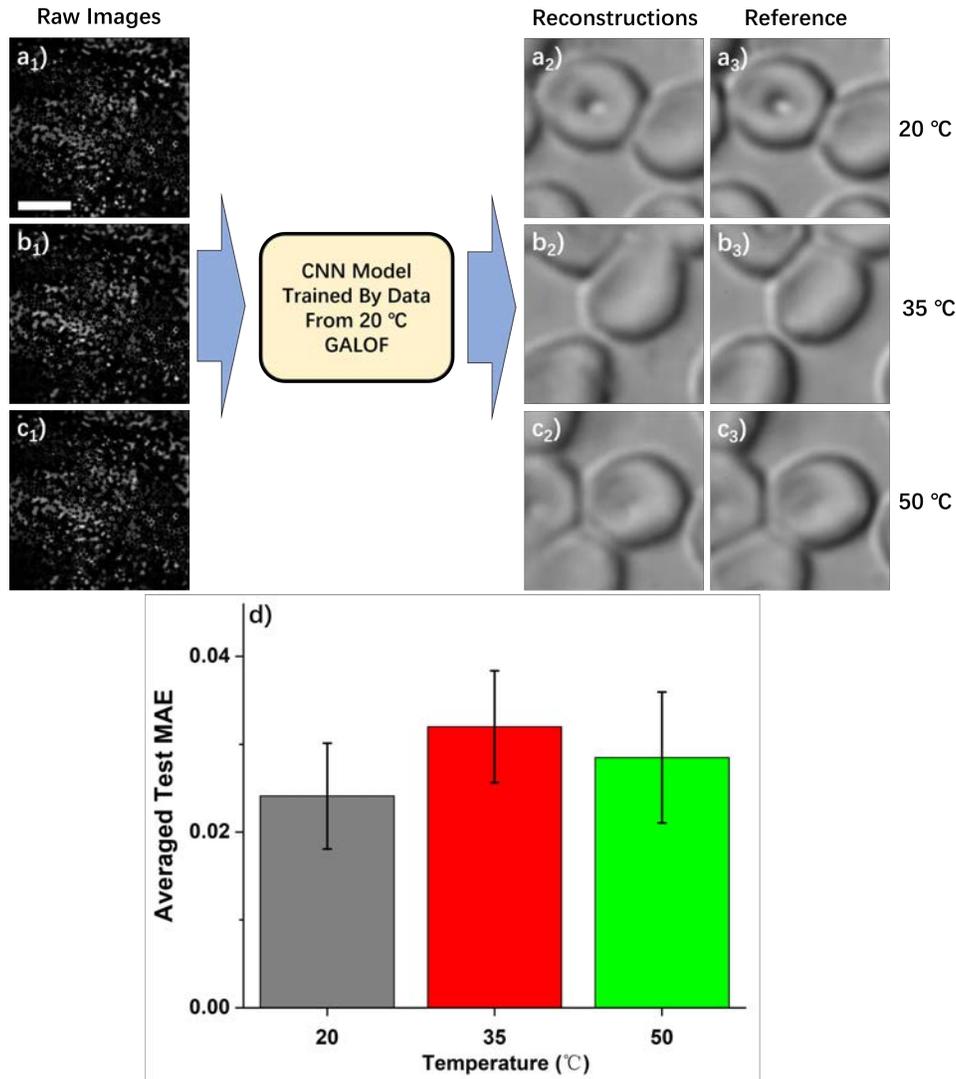

**Figure 5 | Cell imaging at different temperatures. $a_1$)-$c_1$)** are test raw images of human red blood cells collected at 20 °C, 35 °C, and 50 °C, respectively. The length of the scale bar in **$a_1$)** is 4 μm. **$a_2$)-$c_2$)** are the images recovered from **$a_1$)-$c_1$)**. **$a_3$)-$c_3$)** are the corresponding reference images. All data are collected with straight GALOF at 0 mm imaging depth. d) shows the averaged test MAE error for each temperature with the standard deviation as the error bar. More test sample results including reference, raw, and recovered images are provided in Fig. S3 of the supplementary information.

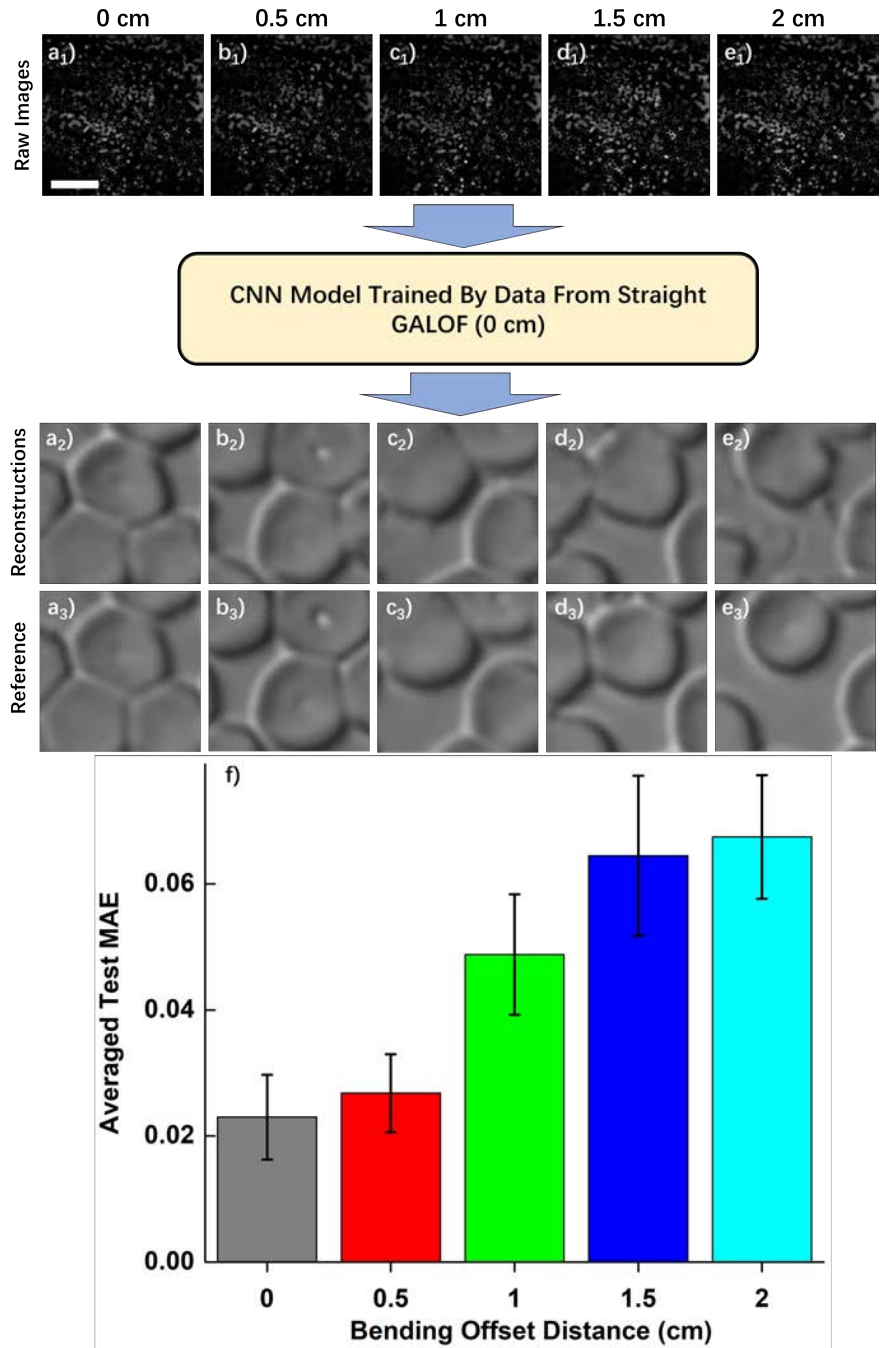

**Figure 6 | Cell imaging under bending.** Data in each column in **a)**-**e)** correspond to examples with the bending offset distance listed above. The definition of offset distance is illustrated in Fig. 1 b). The bending angle range corresponding to offset distances between 0 cm and 2 cm is about 3 degrees, see Methods for more details. **$a_1$)**-**$e_1$)** are raw images collected at different bending offset distances. The length of the scale bar in **$a_1$)** is 4 μm. **$a_2$)**-**$e_2$)** are the images reconstructed from **$a_1$)**-**$e_1$)**. **$a_3$)**-**$e_3$)** are the corresponding reference images. **f)** shows the averaged test MAE for five different bending states with the standard deviation as the error bar. More sample results of human red blood cells including reference, raw, and recovered images are provided in Fig. S4 of the supplementary information.

Next, we test the effect of fibre bending on the performance of our Cell-DCNN-GALOF system. We keep the temperature of the fibre at room temperature and the imaging depth at 0 mm. We collect 15000 image pairs with straight GALOF as the training data and record five sets of separate test data corresponding to five different bending states. Each test set consists of 1000 image pairs. Experimentally, the bending is induced by moving the fibre end by a specified offset distance as illustrated in Fig. 1 b). The relation between the offset distance and the bending angle of the fibre is explained in Methods. We first train the model only using the training data collected from straight GALOF. Then test images from all five different bending states are reconstructed by the non-bending-data trained DCNN model and evaluated using the MAE. The results are shown in Fig. 6. Based on the recovered images in Fig. 6 $a_2$) to $e_2$), high fidelity cell imaging transfer and reconstruction could be performed without any retraining for offset distances smaller than 2 cm (a bending angle of ~3 degrees). The corresponding change of the normalized averaged MAE with bending is depicted in Fig. 6 f). Within this small bending limit every degree of bending results in a MAE increase of about 0.02. This is in sharp contrast to MMF-based systems which require access the distal end of the fibre to recalibrate the transmission matrix if any tiny movement (a few hundred micrometers) of the MMF happens[9,13]. For neuroscience applications[4,13], the flexibility of the Cell-DCNN-GALOF system shows the potential to satisfy the imaging requirements for observing real-time neuron activity in free-behaving objects.

**Cell imaging transfer learning**

We have shown that our DCNN is able to perform high-fidelity image restoration when training and testing is performed with the same types of cells. In practical applications, the Cell-DCNN-GALOF system would be a more efficient and higher functionalized tool if it was able to transfer its learning capability to reconstruct different types of cells which never appeared in the set of training data. To enable transfer-learning reconstruction with high fidelity, a training dataset with high diversity would certainly be beneficial. As a proof-of-concept experiment, we apply a training set with just three different types of images. Sample images are shown in Fig. 7 a)-c). These are images of human red blood cells, frog blood cells, and polymer microspheres. During the recording of data for training, validation, and testing we keep the GALOF straight, the imaging plane at 0 mm depth, and at room temperature. To generate data sets for training and validation, we first collect 10000 image pairs of each human red blood cells, frog blood cells, and polymer microspheres. Subsequently, all 30000 image pairs of three different types are mixed randomly. We extract 28000 image pairs from those randomly mixed images as the training dataset and 1000 image pairs as the validation dataset. To characterize the training process, the accuracy improvement curves during training and validation are tracked and shown in Fig. 7 g). Both curves show convergence to low values after about 20 epochs. The differences between the validation and the training accuracy improvement curves are very small. These characteristics indicate that our DCNN is not overfitting with respect to the training dataset.

As the test data, we record 1000 image pairs from a totally different type of cells, namely bird blood cells. The raw images of the bird blood cells obtained after passing through straight GALOF are shown in Fig. 7 d). These data are fed into the trained DCNN to perform the transfer-learning reconstruction. The reconstructed and reference images are shown in Fig. 7 e) and f), respectively. To enable a quantitative analysis, the averaged test MAE and its standard deviation are provided in Fig. 7 h). A visual inspection demonstrates that within the reconstructed images of bird blood cells one can clearly locate the position and orientation of the nucleus for each single cell. Being trained by a fairly limited set of training data, our DCNN is still able to approximately reconstruct complex cell objects of a totally different type. This transfer-learning capability Cell-DCNN-GALOF system demonstrates that the underlying physics of the imaging process are captured well by the trained DCNN and should prove beneficial for practical applications. At a minimum it can be applied to cell counting tasks in biology and medicine.

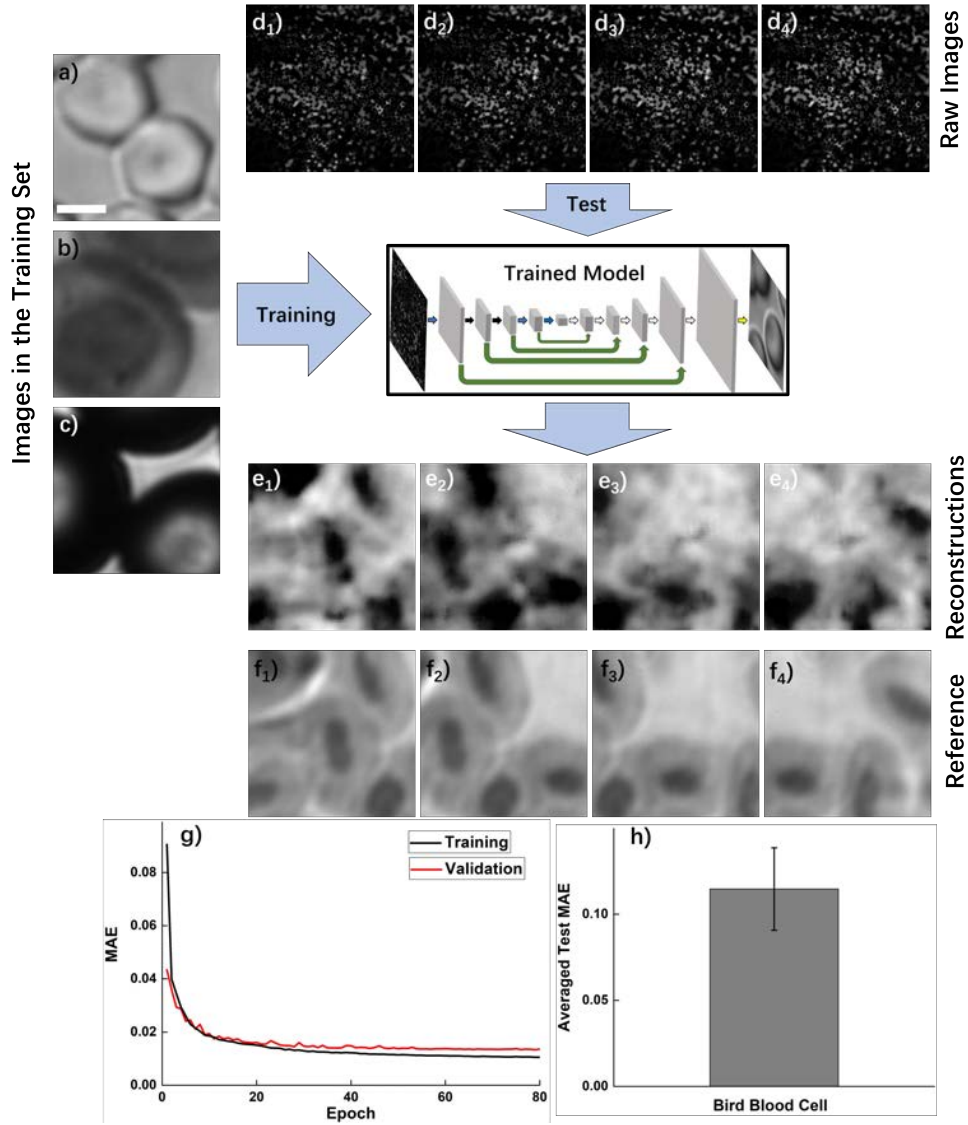

**Figure 7 |Cell imaging transfer learning. a)-c)** are sample cell images in the set of training data. The length of the scale bar in **a)** is 4 μm. There are three different types of cells in the set of training data; **a)** is an image of human red blood cells, **b)** is an image of frog blood cells and **c)** is an image of polymer microspheres. **d)**-**f)** demonstrate the test process using data from images of bird blood cells. **d$_1$)**-**d$_4$)** are the raw images of bird blood cells transported through straight GALOF taken at 0 mm imaging depth and at room temperature. **e$_1$)**-**e$_4$)** are images reconstructed from **d$_1$)**-**d$_4$)**. **f$_1$)**-**f$_4$)** are the corresponding reference images of bird blood cells. **g)** The training and validation accuracy improvement curves using MAE as the metric over 80 epochs. **h)** shows the averaged test MAE of the bird blood cells images with the standard deviation as the error bar.

## Discussion

We have developed a novel Cell-DCNN-GALOF imaging system and demonstrated its ability to provide artifact-free, high fidelity images of different types of cell in real time. Considering the complex morphology of various cells this represents a big leap forward compared to the reported capabilities of previous DCNN based FIOSs imaging systems.

Unlike conventional FOISs using distal optics, DCNN based systems feature the unique ability to image biological objects that are several millimeters away from the fiber facet without any distal optics. This is possible because the DCNN can be trained to accurately model the complete image transfer process including fiber transmission and free space propagation. Despite this advantage, the imaging quality of our Cell-DCNN-GALOF system gradually reduces with increasing imaging depth. This might be attributed to the fact that, under incoherent illumination, high-frequency features of the intensity objects are gradually lost with increasing imaging depth leading to a corresponding rise of the MAE.

Our Cell-DCNN-GALOF system is remarkable robust with respect to thermal and mechanical perturbations of the image transporting GALOF. This qualifies our system particularly for *in vivo* biomedical imaging. This robustness is very hard to be achieved with MMF-based system even if deep learning algorithms are applied. The displayed advantages in performance robustness are based on the different mode propagation properties of GALOFs leading to enhanced single mode guidance and reduced mode interference effects in GALOFs. However, it is worth noting that the imaging quality of the Cell-DCNN-GALOF system starts to degrade if the bending is larger than 3 degree. This might be attributed to remaining multimode channels embedded in the random structure even if most transmission channels demonstrate single-mode properties[34]. The path to improved bending independence will be a further optimization of the random structure inside the GALOF. It should be possible to maximize the scattering in the transverse plane which would further enhance transverse light localization[31,35]. Therefore, GALOF-based imaging system could potentially provide even stronger robustness through optimization of GALOF design and fabrication.

As a proof-of-concept we also show transfer-learning capabilities of our imaging system when images of different cell types that are not part of any training procedure can be reconstructed. We consider these experiments a proof of concept since we believe that a lot of improvement can be achieved in this area when suitable training data with larger diversity are applied. However, generating highly diverse biological training data for a FOIS remains a formidable practical challenge and the computational power available to process large amounts of training data is often an additional bottleneck. We believe that these challenges can be addressed in next generation FIOSs for biological objects with the help of further optimized DCNN architectures.

In conclusion, it is the combination of unique GALOF properties and tailored DCNN design that enables the remarkable capabilities of the presented Cell-DCNN-GALOF imaging system. Both components, GALOF design and DCNN architecture, still have room for improvements and future research will consider both components and their interplay. We are very optimistic that the presented architecture can be the bases for future high-fidelity imaging systems that are minimally invasive and feature robust performance in dynamic environments.

**Methods**

**Fibre fabrication.** The GALOF is fabricated using the stack-and-draw method. Silica capillaries with different diameters and air-filling fractions are fabricated first. The outer diameter of the silica capillaries ranges from about 100 to 180 μm and the ratio of inner diameter to outer diameter ranges from 0.5 to 0.8. To make a preform, capillaries are randomly fed into a silica jacket tube. In the following steps, the preform is drawn to canes with an outer diameter around 3 mm. Finally, the cane is draw to the GALOF with a desired size.

**Experiments and DCNN.** In the experiments of testing the imaging system tolerance with regard to thermal variations, a 10 mm-long section in the middle of the GALOF is heated. To bend the fibre, the input end of the GALOF is fixed while the output end of the GALOF is moved by an offset distance *d*. The relation between the offset distance and the corresponding bending angle of the fibre $\theta$ is given by $d=L[1-cos(\theta)]/\theta$, where *L* is the total length of the GALOF.

The Keras framework is applied to develop the program code for the DCNN. The MAE is defined as $|I_{rec}-I_{ref}|/(wh)$, where $I_{rec}$, $I_{ref}$, *w*, and *h* are the reconstructed image intensity, the reference image intensity, the width, and the height of the images, respectively. The regularization applied in the DCNN is defined by the L2-norm. The parameters of the DCNN are initialized by a truncated normal distribution. For both training and evaluation, the MAE is utilized as the metric. The Adam optimizer is adopted to minimize the loss function. During the training process, the batch size is set at 64 and the training is run through 80 epochs with shuffling at each epoch for all the data shown in this paper. The learning rate is set at 0.005. Both training and test process are run in parallel on two GPUs.

**Data availability**

The data that support the current study are available from the corresponding author under reasonable request.

**Author contributions**

J.Z. designed the convolutional neural network, developed the code for training neural network and image reconstructions, built up the experimental setup, processed the data, and wrote the first draft. J.Z. and J.E.A.L. fabricated the fibre used here. Y.S., H. Z., Z.Z. and S.P. assisted in this work. A.S. and R.A.C conceived the original idea of making the glass-air disordered structure fibre. A.S. led the project and supervised all aspects of the project. All authors discussed the experimental data, revised and approved the manuscript.

**Competing financial interests**

The authors declare no competing financial interests.